\documentclass[letterpaper, 10 pt, conference]{ieeeconf}  

\IEEEoverridecommandlockouts                              

\usepackage{graphicx}
\usepackage{booktabs}
\usepackage{amsmath} 
\usepackage{hyperref}

\title{\LARGE \bf
OpenMines: A Light and Comprehensive Mining Simulation Environment for Truck Dispatching
} 
\author{Shi Meng$^{1}$, Bin Tian$^{2,*}$, Xiaotong Zhang$^{3}$,Shuangying Qi$^{4}$, Caiji Zhang$^{5}$, Qiang Zhang$^{6}$ 
\thanks{*This work was supported by the Key-Area Research and Development Program of Guangdong Province (2020B0909050001), the Natural Science Foundation of Hebei Province (2021402011), the National Key Research and Development Program of China (2022YFB4703700).}
\thanks{$^{1}$Shi Meng is with the National Key Laboratory for Multi-modal Artificial Intelligence Systems, Institute of Automation, Chinese Academy of Sciences, and with the School of Artificial Intelligence, University of Chinese Academy of Sciences, Beijing 100190, China
        {\tt\small mengshi2022@ia.ac.cn}}%
\thanks{$^{2}$Bin Tian is with the National Key Laboratory for Multi-modal Artificial Intelligence Systems, Institute of Automation, Chinese Academy of Sciences, and with the School of Artificial Intelligence, University of Chinese Academy of Sciences, Beijing 100190, China
        {\tt\small bin.tian@ia.ac.cn}}%
\thanks{$^{3}$X. Zhang is with the National Key Laboratory for Multi-modal Artificial Intelligence Systems, Institute of Automation, Chinese Academy of Sciences, and with the School of Artificial Intelligence, University of Chinese Academy of Sciences, Beijing 100190, China
        {\tt\small zhangxiaotong2023@ia.ac.cn}}%
\thanks{$^{4}$Shuangying Qi is with Chongqing Iron and Steel Group Mining Co., Ltd., Chongqing 400050, China
        {\tt\small 303444138@qq.com}}%
\thanks{$^{5}$Caiji Zhang is with the National Key Laboratory for Multi-modal Artificial Intelligence Systems, Institute of Automation, Chinese Academy of Sciences, and with the School of Artificial Intelligence, University of Chinese Academy of Sciences, Beijing 100190, China
        {\tt\small zhangcaiji2021@ia.ac.cn}}%
\thanks{$^{6}$Qiang Zhang is with the Waytous Inc., Qingdao, 266109, China
        {\tt\small qiang.zhang@waytous.com}}%
\thanks{Corresponding author: Bin Tian}%
        }

\begin{document}
\maketitle
\thispagestyle{empty}
\pagestyle{empty}

\begin{abstract}

Mine fleet management algorithms can significantly reduce operational costs and enhance productivity in mining systems. Most current fleet management algorithms are evaluated based on self-implemented or proprietary simulation environments, posing challenges for replication and comparison. This paper models the simulation environment for mine fleet management from a complex systems perspective. Building upon previous work, we introduce probabilistic, user-defined events for random event simulation and implement various evaluation metrics and baselines, effectively reflecting the robustness of fleet management algorithms against unforeseen incidents. We present ``OpenMines'', an open-source framework encompassing the entire process of mine system modeling, algorithm development, and evaluation, facilitating future algorithm comparison and replication in the field. Code
is available in https://github.com/370025263/openmines.

\end{abstract}

\section{INTRODUCTION}

The application of autonomous driving in mining operations, particularly in open-pit mines, represents a significant advancement in the field. The relatively simple road conditions and controllable environment in mines, combined with high labor costs, have made them one of the most promising sub-domains for the implementation of autonomous driving technologies. In open-pit mining, favorable conditions such as good lighting and absence of signal blockages make material transportation an ideal scenario for the early adoption of autonomous driving. As a complex system, unmanned mining operations, where the hauling costs can account for up to 50\% of the total operational costs \cite{moradi_afrapoli_mining_2019}, require a high level of coordination among autonomous trucks, shovels, and mining roads to balance operational costs and output effectively. The Parallel Mining System based on the ACP theory \cite{chen_parallel_2021}\cite{ge_making_2022}\cite{chen_mining_2023}—where A stands for Artificial societies, C for Computational experiments, and P for Parallel execution—is crucial for achieving efficient collaboration in mining fleet operations. This system enables parallel modeling and efficient data synchronization of operational vehicles and mining processes, facilitating complex simulation experiments and parallel execution of multi-agent fleets, thus serving as an effective solution for intelligent fleet management in mining scenarios.

The unmanned mining dispatch system acts as the controller of the mine, adjusting the mine to its highest productivity state by continuously reading mine information and dispatching control commands to subsystems within the mine. The dispatch system of a mine can be divided into production dispatch and extraction dispatch. Production dispatch divides the mine into various mining blocks, aiming to maximize profits over the entire mining lifecycle through a mining plan constrained by transport capacity and excavation costs. This is mainly achieved through solving optimization problems using genetic algorithms and mixed-integer programming, resulting in mining plans. For mine production dispatch, studies have provided common subproblems and datasets suitable for algorithm evaluation \cite{espinoza_minelib_2013}. Mining extraction dispatch mainly involves shovel-truck scheduling, where fleet dispatch is an essential part. In the autonomous mining fleet dispatch, unmanned mining trucks depart from parking areas (charging stations) and request the dispatch algorithm to select the most suitable loading area as the destination. The loading area contains one or more heterogeneous shovels and a parking area serving as a waiting queue. Upon arriving at the loading area, the mining truck first joins the waiting queue in the parking area, then, based on the vehicle dispatch algorithm, selects the optimal electric shovel for loading. After loading, the truck is again dispatched to an appropriate unloading area based on the dispatch algorithm. During transportation, the intersection dispatch algorithm plans the trajectory and speed according to the priority of different vehicles to achieve the best throughput efficiency. The unloading areas can be divided into waste dumps and crushers, accepting different materials. Multiple unloading points in the unloading area require trucks to rejoin the parking area queue and request the vehicle dispatch algorithm again to obtain an optimal unloading position. In both loading and unloading areas, the respective dispatch algorithms are responsible for determining the vehicle's trajectory under given destinations and current traffic conditions to achieve fewer queues and higher output.

The vehicle dispatch algorithm is given real-time information about the current mine state and is essentially solving an optimization problem to provide fleet orders that maximize profits under various emergent conditions. The decision space of the dispatch algorithm is limited, and the problem size is moderate. Even with the current scale of autonomous mining trucks (within 100 vehicles), it is still possible to solve the optimization problem through modeling and heuristic algorithms and mixed-integer programming solvers (such as Gurobi, CPLEX, etc.) in a short time. However, most algorithms do not consider that the mine is a complex system prone to spontaneous emergent events. They calculate the entire vehicle scheduling order at once and re-solve when an emergent event occurs, failing to consider the impact of current decisions on the future state of the system. Reinforcement learning-based vehicle dispatch algorithms \cite{huo_reinforcement_2023}\cite{zhang2023vehicle} view the provision of vehicle orders as a sequential decision-making problem over time. They train an optimal strategy, not just a solution, through custom reward functions and curriculum training, representing a promising direction. However, current implementations of reinforcement learning-based vehicle dispatch algorithms have not incorporated uncertainty and emergent events into their simulation environments, failing to fully demonstrate the potential of reinforcement learning methods.

This paper makes several contributions: 1) It categorizes and summarizes current mine vehicle dispatching algorithms in the context of mining scenarios. 2) To address the issue of disparate, proprietary implementations hindering uniform evaluation and comparison of dispatch algorithms, we develop an open-source simulation environment for mine dispatching. 3) We replicate representative dispatching algorithms and compare them within a unified evaluation framework. 

\section{Related Work}
In mining scenarios, the dispatch problem can be succinctly described as assigning suitable destinations to each mining vehicle based on the current state of the mine to maximize productivity or other objectives. There are two main solutions to this problem: optimization modeling and reinforcement learning. In optimization approaches, an optimal goal (usually a combination of production, cost, and energy consumption) is defined manually. Combined with constraints like fleet scheduling, vehicle fuel consumption, and mileage, the best vehicle scheduling plan is sought or solved within feasible domains, presenting an NP-hard problem.

Given the complex nature of mining systems, when random events like traffic jams, vehicle breakdowns, or excavator malfunctions occur, pre-determined optimal solutions become unfeasible and need to be recalculated. In some cases, random events cause the system to exceed the bounds of optimal modeling, and the resulting solutions can be counterproductive. In such situations, manual dispatching and other methods are required to increase the flexibility of algorithms. The study \cite{seiler_flow-achieving_2022} utilized a flow-achieving scheduling tree (FAST) integrated with Monte Carlo tree search (MCTS) to enhance algorithm resilience and consider the subsequent impacts of current decisions. Reinforcement learning methods do not start by directly generating all dispatch orders. Instead, they view the vehicle target assignment problem as a sequential decision-making strategy. In the event of random occurrences, reinforcement learning methods \cite{zhang_dynamic_2020} are more capable of making timely and rational target assignments. However, the interpretability of reinforcement learning methods is poor. When trained assignment strategies cause issues, it is difficult to attribute and correct them promptly. Additionally, the simulation environment for reinforcement learning needs to be sufficiently complex to ensure good generalization in real mine dispatch scenarios, which is currently lacking.

Current research on dispatch simulation environments is mostly not open-source and has certain modeling deficiencies. Studies \cite{zhang_scheduling_2021} and \cite{zhang_real-time_2022} used real data to calibrate mining models for simulation environments, modeling the relationship between truck energy consumption and target orders effectively. However, these simulation environments do not cover random events and actual mine traffic simulation. The decision-making process does not consider the congestion level of roads. \cite{seiler_flow-achieving_2022} divided real mine paths into RoadSegments, and accumulated historical data from actual truck operations for each segment to simulate the actual running time of mining trucks in the environment, also considering the movement of electric shovels during production scheduling. However, this method still does not approximate real traffic conditions, and current order decisions do not affect road throughput times. Additionally, random events like weather, vehicle wear, and tear are not considered. \cite{huo_reinforcement_2023} established a reinforcement learning simulation environment, encapsulated with OpenAI interfaces, effectively modeling vehicle emissions. \cite{zhang_dynamic_2020} used discrete simulation components like SimPy for reinforcement learning environments, sampling historical operation data using a gamma distribution to determine vehicle arrival times, yet still did not consider the impact of random events and current decisions on subsequent traffic. \cite{both_joint_2020} and \cite{torkamani_linkage_2015} considered production scheduling and fleet dispatch together, accounting for vehicle movement between different mining pits. \cite{mohtasham_optimization_2021} proposed a Chance-Constrained Goal Programming (CCGP) model based on four important objectives to estimate the impact of uncertainty on the efficiency of the truck-shovel system, explicitly modeling the randomness of the mining system in order generation. Research \cite{moradi_afrapoli_multiple_2019} used commercial simulation systems like Modular Mining DISPATCH for fleet simulation and modeled real-time dispatch of mining trucks with a multi-objective transportation model. As mining scenarios are commercial, the dispatch algorithm works mentioned above are not open-source and have independently implemented their simulation environments, making it difficult to replicate and compare different algorithms.

In summary, the current simulation environments for mine dispatch algorithms mainly face the following issues: 1) Lack of consideration for emergent situations encountered by vehicles and mine systems in actual operations. 2) The simulation environment does not consider the impact of current decisions on future states. 3) Traffic in the mine is not simulated but rather sampled randomly using historical data. 4) The code is not open-source, and independent implementation of simulation environments poses challenges for replication and comparison. We addressed these issues by designing and developing the OpenMines framework for evaluating and improving mine fleet dispatch algorithms. By thoroughly considering existing simulation environment problems, we introduced random events and traffic simulation to enhance the robustness and practical performance of algorithms.

\section{OpenMines}
\subsection{Rethinking Dispatch Algorithms of Mining Trucks}
Given the current state $S_0$ of the mining fleet, operations research-based algorithms provide an optimal solution $a_0$ with respect to $S_0$. However, should an incident occur, forcing the system to evolve to state $S_1$, OR methods will have to recalculate and offer $a_1$ as the optimal solution with respect to the new state. If we consider the system as time-invariant, OR methods solve the problem well. However, the mining fleet exhibits time-variant and stochastic nature, making $a_0$ and $a_1$ need to be viewed on the same timeline and merged as $a_{01}$, which is likely not the optimal solution. OR methods have overlooked the time-variant nature of the mining system. The reinforcement learning approach models the problem well by treating the dispatch problem as a time-related decision problem. However, most RL approaches haven't introduced sufficient time-variance into the simulation system.

Mining truck dispatch algorithms fundamentally act as controllers for the complex mining system. Such systems exhibit characteristics like non-linearity, emergence, self-organization, adaptability, and feedback loops. Therefore, dispatch algorithms must be designed with full consideration of potential changes in the mining environment and the impact of their own historical decisions on the mining system.

\subsection{Module Introspection}
\begin{figure}[t]
  \centering
   \includegraphics[width=1\linewidth]{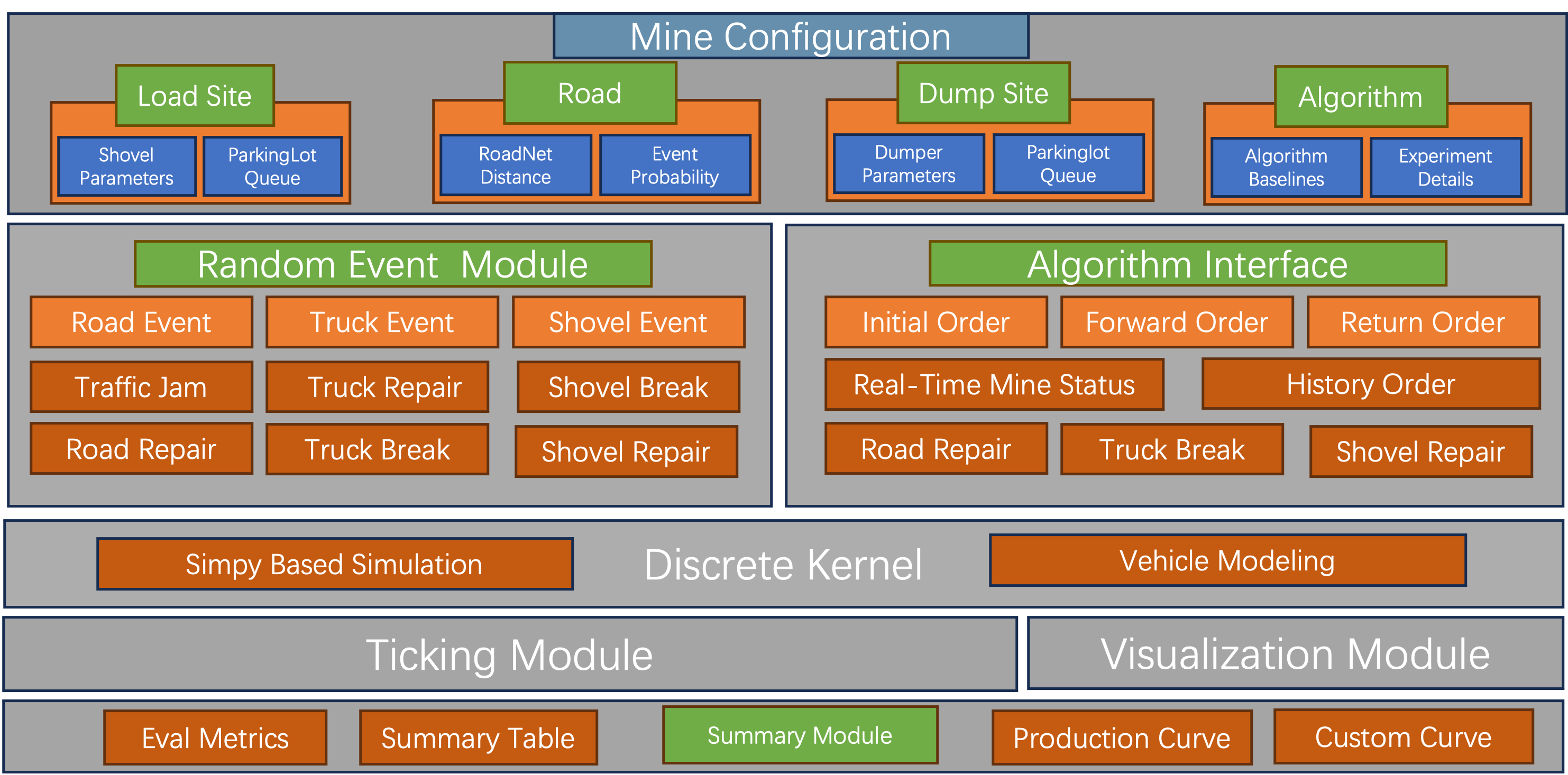}
   \caption{The OpenMines architecture, including configuration, random event and algorithm interface details.}
   \label{fig:arch}
\end{figure}

OpenMines divides the mining system into a collection of subsystems(Fig. \ref{fig:arch}), including charging areas, loading zones, unloading zones, and roads. The core of the system is a discrete event simulation engine based on SimPy. During fleet operation, random events occur through probabilistic sampling, and the decisions of dispatch algorithms influence the distribution of these events. To achieve lightweight but effective event simulation, we have not used traffic frameworks like SUMO for true modeling of mining transport scenarios but instead adopted a method of human-defined events to simulate traffic.

\subsubsection{Discrete Kernel}
The discrete kernel uses an object-oriented design pattern, modeling unloading zones, loading zones, and charging areas. The loading zone includes multiple heterogeneous shovel fleets configurable via a file and parking areas for waiting and statistics. The unloading zone contains unloading spots and parking areas for queuing and statistics. The modeling of trucks is the core of the discrete kernel, employing an event queue to track and record their operational states, including six basic states: "waiting for loading, loading, fully loaded, waiting for unloading, unloading, and empty run." During operation, there is a probability of trucks randomly malfunctioning and going offline, completely changing the parameters of the mining fleet dispatch problem.

\subsubsection{Random Event Module}
The random event module has two main functions: simulating traffic situations and non-traffic random events. It records the number of vehicles and journey completion on each road during simulation, with traffic conditions manifested through the occurrence of congestion events, following the approximation in Eq. (\ref{eq:jam_equ}).
\begin{equation}\label{eq:jam_equ}
f(t,c) = \frac{\sum_{i} C_i(t) \cdot N(\mu,\sigma) }{\sum_{i} C_i(t)}
\end{equation}
Here, \(t\) represents the simulation time, and \(C_i(t)\) represents the journey completion rate of truck \(i\) at time \(t\). We consider the probability of each vehicle encountering a traffic jam to be the sum of normal distributions of nearby vehicles. Each time a truck departs on its journey, the random event module samples potential jam locations based on distribution in Eq. (\ref{eq:jam_equ}), then determines the duration of the jam using a Weibull distribution and compares it with the Truck Arrival Time to assess if the current truck journey will be affected by the jam.
The random event module also models the availability of roads. We consider road maintenance to follow an exponential distribution. At each unit of time, the road samples the outcome from distribution in Eq. (\ref{eq:road_repair}). When the sampled time is less than the current time \(t\), a maintenance event occurs, and vehicles passing the maintained road receive a percentage time penalty through normal distribution sampling.
\begin{equation}\label{eq:road_repair}
f(t) = \lambda e^{-\lambda t}; (t>0)
\end{equation}
We use a similar approach to Eq. (\ref{eq:road_repair}) for simulating the availability of mining trucks. When a maintenance event occurs for a mining truck, it will stop moving and wait for repair, with the repair duration satisfying a normal distribution. When a truck damage event occurs, the vehicle will return to the charging area and no longer accept dispatch algorithm instructions.
We use a similar approach to Eq. (\ref{eq:road_repair}) for simulating the availability of shovels. When a shovel maintenance event occurs, trucks already in the queue continue to wait, and the shovel itself no longer accepts new orders. Its repair duration satisfies a normal distribution. When a shovel damage event occurs, the shovel no longer accepts dispatch algorithm instructions, and the trucks in queue will request dispatch again.
\subsubsection{Ticking Module}
Rapid iteration and visualization are essential for problem discovery in the simulation environment, with monitoring during operation being crucial. Each subsystem in the mining system, such as roads, trucks, shovels, unloading zones, and loading zones, saves its operational status and statistical information in the form of tick files as simulation output. We store the information of the mining operation process in an event pool and frame-by-frame analyze the condition of vehicles at each unit event post-simulation.
\subsubsection{Visualization Module}
\begin{figure}[t]
  \centering
   \includegraphics[width=1\linewidth]{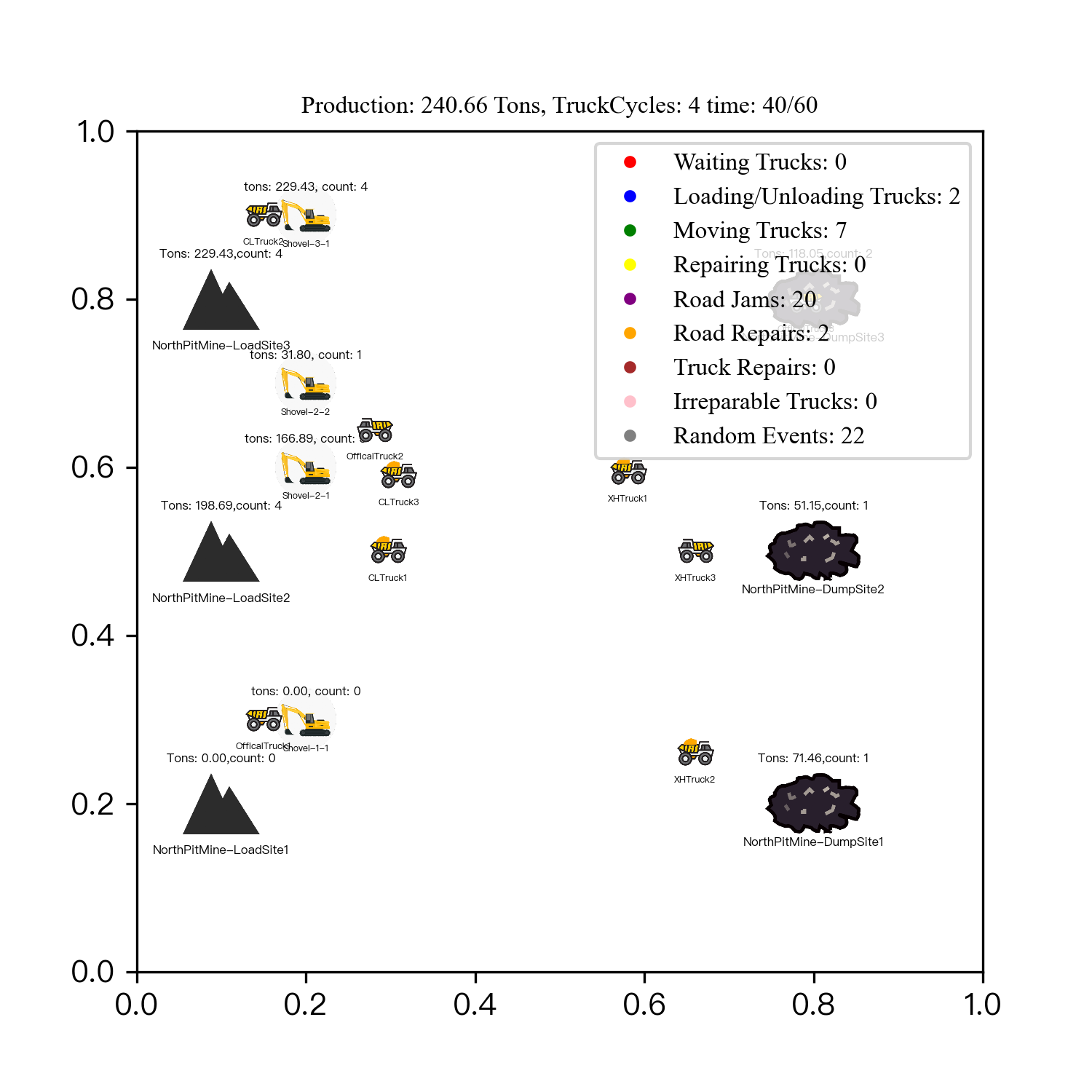}
   \caption{The OpenMines visualization module offers real-time visualization capabilities for assessing the performance of truck dispatching algorithms. This image is a snapshot from the generated GIF.}
   \label{fig:vis}
\end{figure}

Visualization is an important error correction step. Here, we use Matplotlib combined with truck journeys to draw a bird's-eye view of the mining operation like in Fig. \ref{fig:vis}. The module takes tick files as input, analyzes vehicle statuses frame by frame, and outputs corresponding GIFs. The visualization module distinguishes between the states of mining trucks such as initialization, empty load, full load, and waiting, and also reflects the shovel queuing scenario. A video demonstration of the visualization module is available online.\footnote{\url{https://figshare.com/articles/media/Untitled_Item/25124165}}
\subsubsection{Logging Module}
The discrete kernel also records logs for each run, assisting in analyzing problems within the custom simulation environment. However, most performance issues of dispatch algorithms can be identified through GIFs from the visualization module. Log analysis is only necessary when performing quantitative attribution of specific features of the dispatch algorithm.
\subsubsection{Summary Module}
The summary module summarizes indicators such as Matching Factor \cite{burt_match_2007}, average decision time, and mine output. After each simulation run, it generates a performance report for the current algorithm and compares it with algorithms already in the configuration file. We will showcase these in the KPI Section.

\subsection{Algorithm Key Indicators}
\subsubsection{Algorithm Decision Latency}
Algorithm Decision Latency (ADL) represents the average time taken by the algorithm to make each effective decision. The time consumed by policy-based and reinforcement learning algorithms is significantly less than that required by optimization-based algorithms, which often need several seconds to re-solve with each change. In contrast, policy and reinforcement learning methods have consistent decision times, with each decision taking milliseconds. The ADL of our framework is calculated according to the following Eq. (\ref{eq:adl}).

\begin{equation}\label{eq:adl}
f(t,c) = \frac{\sum_{j=1}^{M} T_0 + \sum_{i=1}^{N} t_i}{N}
\end{equation}
Where \(T_0\) represents the initialization time of the algorithm, for optimization-based methods, \(T_0\) is the time taken to compute the entire plan. \(M\) represents the number of times the algorithm is triggered for initialization. \(N\) is the number of dispatch orders. \(t_i\) is the time taken for generating each individual order. When the average decision time of the algorithm is too long, many vehicles will wait for dispatch from the central system, thereby slowing down the overall system efficiency.

\subsubsection{Match Factor}
The Match Factor is the most representative metric for evaluating algorithms in the domain of mine fleet dispatch. We have implemented the Match Factor for heterogeneous shovel and truck fleets as per \cite{burt_match_2007}.The Match Factor (MF) is defined as follows:

\begin{equation}\label{eq:mf}
MF=\frac{ N * \sum_{i} lcm(ULT)_i}{  \sum_{j}\frac{shovel_j*lcm(ULT)_i}{ULT_{i,j}} (truckCycleTime)} 
\end{equation}
Where \(N\) represents the number of trucks, \(i\) represents the type of truck, \(j\) represents the type of shovel, \(ULT\) represents the unique loading time of shovel. \(lcm(ULT)_i\) is the least common multiple of the time required for all types of shovels to fill trucks of type \(i\). \(ULT_{i,j}\) is the time taken for trucks of type \(i\) to be filled by shovels of type \(j\). \(shovel_j\) is the number of shovels for a specific shovel type. \(truckCycleTime\) is the average time taken for the entire fleet to complete one cycle of waiting for loading, loading, full load, waiting for unloading, unloading, and empty load.

When the value approaches 1, it indicates that the fleet is in an optimal state, achieving a balance in efficiency and ratio. An MF value greater than 1 suggests redundancy in the number of mining trucks; a value less than 1 implies either shovel redundancy or wastefulness in the dispatch algorithm, causing many shovels to idle. However, a value slightly below 1 typically indicates lower overall transportation costs.

\subsubsection{Production}
Production is defined as the total tonnage collected by all unloading areas. It is represented as a discrete, monotonically increasing curve throughout the operation, with extended horizontal segments on the curve indicating prolonged idle times in the system.

\subsubsection{Total Wait Time}
Total wait time represents the time vehicles spend queuing in loading and unloading areas. Excessive wait times within the same simulation duration reflect inefficiencies in the mining system. Ideally, mining trucks should have zero wait time, allowing immediate execution of loading and unloading upon arrival at destinations.

\subsubsection{Custom Indicators}
Other custom indicators can be expanded into the KPI module. OpenMines supports custom indicators.

\section{Benchmarks}
\subsection{Policy Baselines}
Policy baselines are relatively simple dispatch methods following basic rules like nearest neighbor, random targets, fixed grouping, etc. This study implements common policy baselines. When random events occur, policy-based baselines cannot respond specifically, as implemented and referenced from \cite{zhang_dynamic_2020}.

\subsubsection{Random Policy}
The random policy randomly determines the vehicle's destination, following a uniform distribution $P\{X\} = \frac{1}{N}$. This policy can distribute vehicles evenly across the network but struggles to utilize high-capacity areas effectively.

\subsubsection{Nearest Policy}
The nearest policy is simple, following a greedy strategy based on spatial distance for each decision, selecting the shortest path. However, this can lead to vehicle congestion in closer sections, wasting transport capacity.

\subsubsection{Shortest Queue (SQ) Policy}
As per \cite{subtil_practical_2011} and \cite{zhang_dynamic_2020}, at the initialization of mining trucks, the dispatch algorithm randomly selects all loading points. Once a truck reaches a loading point, the strategy assigns it to the shortest queue destination. Trucks already on the road are considered in the queue ahead.

\subsubsection{Shortest Processing Time First (SPTF) Policy}
This strategy \cite{rose_shortest_2001} selects the loading/unloading area with the shortest processing time for the current vehicle, where processing time includes expected waiting time in the queue and loading time. Vehicles on the road are also accounted for in the waiting queue.

\subsection{Fixed Group Method}
In practice, a fixed grouping strategy is often used, binding a certain number of vehicles to specific shovels for transport. This stems from different transport contractors in mines, usually with individual dispatchers managing emergencies and route assignments.

This study implements a rule-of-thumb grouping strategy. We first calculate the output rate of loading areas following Eq. (\ref{eq:productivityRatio}), then group trucks of different capacities into loading areas until the output matches the fleet's capacity. In loading areas, trucks are assigned to shovels based on the shortest queue principle. For selecting unloading areas, trucks use the nearest distance principle.

\begin{equation}\label{eq:productivityRatio}
\text{Productivity Ratio}_j = \frac{\text{shovelNum}_j \cdot \frac{\text{shovelSize}_j}{\text{shovelTime}_j}}{\sum_i \text{shovelNum}_i \cdot \frac{\text{shovelSize}_i}{\text{shovelTime}_i}} 
\end{equation}
Where \(j\) represents the index of the loading area.

\subsection{Comparison of Algorithm Performance}
\begin{figure}[htbp]
  \centering
   \includegraphics[width=0.5\textwidth]{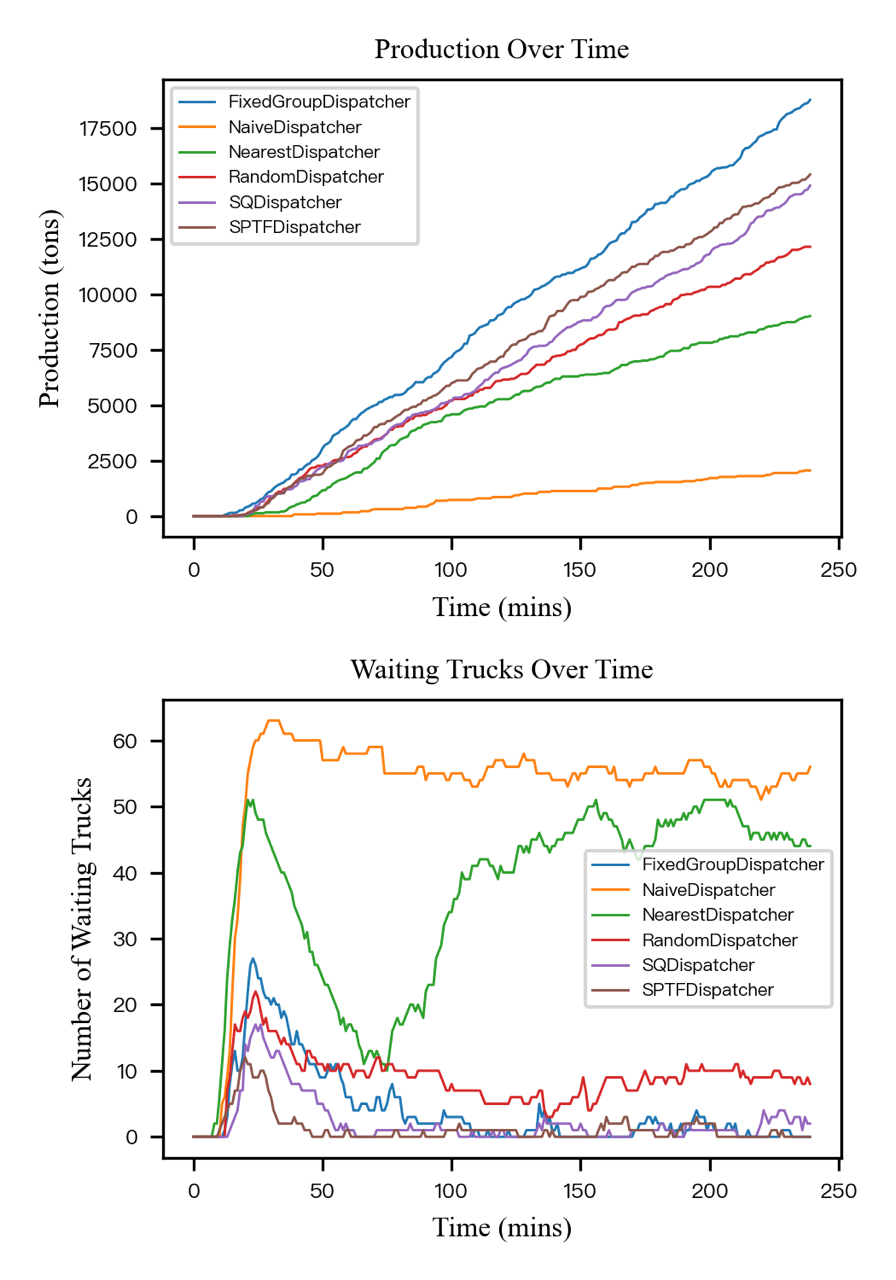}
   \caption{The production and waiting truck curve of baselines.}
   \label{fig:curve_policy}
\end{figure}

We referenced the operation information of Huolinhe Open-pit Coal Mine in September 2022, desensitized the relevant data, and configured the OpenMines environment. In the experiment, we simulated over 4 hours with a unit time of 1 minute, for three different fleets totaling 71 trucks and 21 heterogeneous shovels across five loading and unloading points. The result is displayed in Fig. \ref{fig:curve_policy} and Table \ref{tab:performance_metrics}.

\begin{table}[htbp]
\centering
\begin{tabular}{@{}lcccc@{}}
\toprule
\textbf{Name} & \textbf{Produced} & \textbf{Matching} & \textbf{Total Wait} & \textbf{Road} \\
 & \textbf{Tons} & \textbf{Factor} & \textbf{Time} & \textbf{Jams} \\
\midrule
FixedGroupDispatcher & 14909.56  & 1.27 & 308.94 & 605 \\
NaiveDispatcher & 2266.54  & 0.51 & 9547.48 & 120 \\
NearestDispatcher & 6582.21  & 0.48 & 4061.66 & 297 \\
RandomDispatcher & 10627.08  & 0.82 & 484.75 & 407 \\
SQDispatcher & 13232.29  & 1.03 & 314.80 & 499 \\
SPTFDispatcher & 13096.42  & 0.95 & 225.25 & 492 \\
\bottomrule
\end{tabular}
\caption{Dispatcher performance result.}
\label{tab:performance_metrics}
\end{table}

According to Fig. \ref{fig:curve_policy}, policy baselines such as the Naive and Nearest Dispatch strategies perform similarly and underperform compared to the random baseline. The Naive strategy, with no specific dispatching, leads to most vehicles queuing rather than hauling, resulting in the fewest traffic jam events. The Nearest Dispatch strategy reduces road time and somewhat disperses vehicles across different loading areas, thereby decreasing total cycle time and increasing output and reduced wait times. The Random strategy, by uniformly dispersing vehicles, achieves the lowest frequency of traffic jams among the last three strategies but spends more time on longer routes due to the lack of consideration of distances between mining areas. The SPTF and SQ methods, by considering loading time on the road rather than just numbers, achieve shorter wait times and fewer traffic jams. The Fixed Group method effectively distributes fleets across loading points, achieving the best output but is prone to traffic jams due to its inability to adapt strategies on-site, resulting in relatively long wait times.

\section{Conclusions}
Mining truck dispatch algorithms are crucial for efficient mining operations. However, most current research is based on proprietary projects with non-open-source code, posing challenges for replication and comparison. Many simulation environments overlook the modeling of random events such as traffic, mining trucks, and shovels, crucial for algorithm robustness and practical performance. We summarized the main works in mine fleet management, analyzed and evaluated their simulation environments, and introduced random events and traffic simulation into the environment. Representative baseline strategies were implemented for benchmarking and comparative analysis. The open-source environment OpenMines, based on discrete event simulation and random event modeling, was developed for algorithm comparison and problem visualization in future works. Notably, we implemented a large-language-model-based dispatch strategy in the framework, capable of responding well to emergencies and accepting human natural language instructions, potentially serving as an effective decision-support tool for dispatchers.

\addtolength{\textheight}{-12cm}   







\bibliographystyle{./IEEEtran} 
\bibliography{./IEEEabrv,./IEEEexample}

\end{document}